# Anderson localization of spinons in a spin-1/2 antiferromagnetic Heisenberg chain

B. Y. Pan, S. Y. Zhou, X. C. Hong, X. Qiu, & S. Y. Li[*]

*Department of Physics, State Key Laboratory of Surface Physics, and Laboratory of Advanced Materials, Fudan University, Shanghai 200433, China*

**Anderson localization is a general phenomenon of wave physics, which stems from the interference between multiple scattering paths[1,2]. It was originally proposed for electrons in a crystal, but later was also observed for light[3-5], microwaves[6], ultrasound[7,8], and ultracold atoms[9-12]. Actually, in a crystal, besides electrons there may exist other quasiparticles such as magnons and spinons. However the search for Anderson localization of these magnetic excitations is rare so far. Here we report the first observation of spinon localization in copper benzoate, an ideal compound of spin-1/2 antiferromagnetic Heisenberg chain, by ultra-low-temperature specific heat and thermal conductivity measurements. We find that while the spinon specific heat $C_s$ displays linear temperature dependence down to 50 mK, the spinons thermal conductivity $\kappa_s$ only manifests the linear temperature dependence down to 300 mK. Below 300 mK, $\kappa_s/T$ decreases rapidly and vanishes at about 100 mK, which is a clear evidence for Anderson localization. Our finding opens a new window for studying such a fundamental phenomenon in condensed matter physics.**

Since its discovery[1], Anderson localization has been one of the most studied phenomena in physics[2-13]. Originally, it was found that beyond a critical amount of impurity scattering, the electrons in a crystal are trapped and the conductivity vanishes[1]. This counterintuitive result stems deeply from the quantum interference effect of electron waves[2]. Later, Anderson localization was widely observed for light[3-5], microwaves[6], ultrasound[7,8], and recently for ultracold atoms[9-12], thus it becomes a general phenomenon of wave physics.

Back to the crystal, if some atoms have magnetic moment, there may exist magnetic excitations such as magnons and spinons. We aim to look for Anderson localization of these magnetic quasiparticles, which has been rarely studied. Since one-dimensional (1D) systems provide a paradigmatic example of strong Anderson localization[13], it is natural to focus on the quasi-1D spin chains, such as spin-1/2 Ising chain or antiferromagnetic Heisenberg chain (AFHC).

In zero magnetic field, the domain-wall quasiparicles of a spin-1/2 Ising chain are gapped[14]. On the contrary, the spin-1/2 isotropic AFHC is described by the Hamiltonian

$$H = J\sum_i S_i S_{i+1},$$

representing the antiferromagnetic interactions between nearest neighbour spins $S_i$ along the 1D chain with coupling strength $J$. It is well-known that the magnetic excitations in the chain are gapless spinons[15]. Figure 1a shows a carton picture of those spinons, and Fig. 1b sketches the "two-spinon continuum" spectrum, limited by the lower bound $\varepsilon_1 = \frac{\pi}{2}J|\sin q|$ and upper bound $\varepsilon_2 = \pi J |\sin q/2|$. Therefore it is more suitable to study Anderson localization of spinons in the compounds of spin-1/2 AFHC.

The transport properties of the spinons in a spin-1/2 AFHC have been studied by thermal conductivity measurements for various compounds[16,17]. The intrachain coupling $J/k_B$ ranges from 2000 K in $Sr_2CuO_3$ (ref. 18) to 10 K in $Cu(C_4H_4N_2)(NO_3)_2$ (ref. 19). However, most of those compounds are not ideal AFHC at low temperature, due to the existence of weak interchain coupling $J'$. For example, $Sr_2CuO_3$ shows three-dimensional magnetic order below $T_N = 5.4$ K (ref. 20). Furthermore, all the measurements were limited above 0.3 K. So far, the spinon Anderson localization has not been observed.

Copper benzoate, $Cu(C_6H_5COO)_2 \cdot 3H_2O$, is a rare ideal compound of the spin-1/2 AFHC[21]. It has a monoclinic crystal structure with lattice constants $a = 6.98$ Å, $b = 34.12$ Å, $c = 6.30$ Å, and $\beta = 89.5°$ (ref. 22), as shown in Fig. 1c. The $Cu^{2+}$ ions carry spin-1/2 and form chains along the $c$ axis. With the coupling $J/k_B \approx 18.6$ K, no magnetic Braggs peaks were found at $T = 0.3$ K and no specific heat anomaly associated with magnetic phase transition was observed down to $T = 0.1$ K (ref. 23), indicating the interchain coupling $J'$ is extremely weak.

Figure 2a shows the specific heat of copper benzoate single crystal in zero magnetic field from 3.5 K down to 50 mK. The curve can be well fitted by $C_{total} = aT + bT^3$, giving $a = 0.291 \pm 0.001$ J/mol K$^2$ and $b = 0.0094 \pm 0.0001$ J/mol K$^4$. Since copper benzoate is an insulator, the first linear term $aT$ is apparently contributed by the spinons, due to their low-energy linear dispersion shown in Fig. 1b. For spin-1/2 AFHC compounds, the spinon specific heat at $T \ll J/k_B$ can be theoretically calculated by the formula $C_s = \frac{2Nk_B^2}{3J}T$, where $N$ is the number of magnetic ions per unit volume (ref. 24-26). With $N = 2.67 \times 10^{27}$ m$^{-3}$ and $J/k_B = 18.6$ K for copper benzoate, one obtains $C_s/T = 0.298$ J/mol K$^2$, which is in good agreement with our experimental value 0.291 J/mol K$^2$. The second term $bT^3$ comes from the phonons. In Fig. 2b, we plot the magnetic specific heat of copper benzoate in $H = 0$ and 7 T below 1 K. The small phonon contribution has been subtracted in both cases. Similar magnetic specific heat data have been previously obtained down to 100 mK by Dender *et al.*[23].

From Fig. 2b, the magnetic specific heat in $H = 7$ T shows a clear signature of field-induced energy gap. The field dependence of this gap was found to obey $\Delta(H) \propto H^{2/3}$ (ref. 23). This has been understood on the basis of quantum sine-Gordon model, by considering the stagged field $h$ induced by the external field $H$ (ref. 27-29). According to this model, the magnetic excitations of copper benzoate in magnetic field are solitons, antisolitons, and breathers. In Fig. 2b, the magnetic specific heat in $H = 7$ T can be fitted in the framework of the sine-Gordon theory below 0.7 K with a gap of 2.41 K (ref. 30), which suggests it is indeed contributed by those magnetic excitations.

We now turn to the thermal conductivity measurements, which explore the transport properties of those magnetic excitations in copper benzoate. Figure 3a presents the thermal conductivity data of sample S1 in $H = 0$, 7, and 14.5 T. In zero field, the data between 0.3 and 0.7 K can be well fitted by $\kappa = aT + bT^{\alpha}$, giving $a = 1.70\pm0.04$ mW/cm K$^2$, $b = 8.87\pm0.04$, and $\alpha = 2.80\pm0.03$. The term $bT^{\alpha}$ is a typical contribution of phonons. For phonon scattering off the crystal boundary at low temperature, one usually gets $\alpha = 3$, but specular reflection of phonons at the smooth crystal surfaces can result in a lower power $\alpha < 3$ (ref. 31).

Usually electrons will contribute a linear term of thermal conductivity in a metal. Again, since copper benzoate is an insulator, we attribute the linear term $aT$ in $H = 0$ T to the spinons. The same linear temperature dependence of spinon specific heat and thermal conductivity in $H = 0$ T suggests that these two quantities are directly related by the simple kinetic expression $\kappa_s = C_s v_s l_s$, with constant spinon velocity $v_s$ and mean free path $l_s$. The $v_s$ can be calculated via $v_s = Jd\pi/2\hbar$, where $d$ is the distance between the spins along the chain direction. (ref. 32). For copper benzoate, we get $v_s \approx 1.2 \times 10^3$ m/s.

With the experimental data of $\kappa_s$ and $C_s$, $l_s \approx 1060$ Å is further obtained, corresponding to about 330 spin distances along the chain direction. Previously, the linear temperature dependence of $\kappa_s$ was only observed in the range 100-300 K for the pseudo-two-leg ladder compound $CaCu_2O_3$ ($J/k_B \approx 2000$ K and $T_N \approx 25$ K), where the small $l_s \approx 22$ Å is due to the high density of static scattering centers[33].

In $H = 7$ T, the magnetic excitations of solitons, antisolitons, and breathers will replace spinons, according to the quantum sine-Gordon model. After subtracting the phonon contribution $bT^\alpha$, the magnetic thermal conductivity $\kappa_s$ in $H = 7$ T is plotted in Fig. 3b together with $C_s$ (scaled by 6 times). One can see that $\kappa_s$ shows nearly the same temperature dependence as $C_s$, indicating that $\kappa_s$ in $H = 7$ T indeed comes from those solitons, antisolitons, and breathers. Further increasing field will result in larger energy gap, thus the thermal conductivity in $H = 14.5$ T should further approach the phonon background, as observed in Fig. 3a. These consistent behaviours of thermal conductivity in magnetic field also support that the linear term $aT$ in $H = 0$ T is contributed by spinons.

By measuring more single crystals of copper benzoate, we find that the residual linear term $\kappa_s = aT$ is sample dependent. In Fig. 3c and 3d, the $\kappa/T$ of samples S2 and S3 in $H = 0$ T are plotted versus $T^{1.90}$ and $T^{1.85}$, respectively. While the phonon contributions are similar to S1, their $\kappa/T$ extrapolate to different values of $\kappa_s/T$, 0.60±0.02 and 1.25±0.05 mW/cm K$^2$ for S2 and S3, respectively. For clarity, we subtract the phonon contributions and plot $\kappa_s/T$ for all three samples in Fig. 4, which shows that the $\kappa_s/T$ can differ by 3 times. This suggests that the spinon mean free path in copper benzoate is sample dependent, which is reasonable since different sample may have different disorder level. A good analogy of above spinon thermal conductivity in the spin-1/2 AFHC is the electron thermal conductivity in the metal.

More interesting observation from Fig. 4 is that $\kappa_s/T$ decreases rapidly below 300 mK, and vanishes at about 100 mK. Previously, a low-temperature downturn of thermal conductivity has been observed in cuprate superconductors, which is attributed to the thermal decoupling of electrons and phonons in the samples[34,35]. Here we interpret that the vanishing $\kappa_s/T$ below 300 mK in copper benzoate is not due to the spinon-phonon decoupling, but rather the manifestation of Anderson localization. First, as seen in Fig. 2, the linear term of spinon specific heat can be measured nicely down to 50 mK, showing the good thermal coupling between the spinons and phonons in the sample. Secondly, 1D system is the best place for the happening of Anderson localization, therefore it is actually not very surprising that we finally observe the spinon Anderson localization in this ideal spin-1/2 AFHC compound.

For Fig. 4, the onset temperature of the spinon localization $T_{AL}$ remains at about 300 mK when the mean free path of spinons differs by 3 times between S1 and S2. This indicates that $T_{AL}$ is not very sensitive to the disorder level. A numerical simulation on the effect of disorder on the spinon localization in spin-1/2 AFHC[36], specified to copper benzoate, will be very helpful.

In conclusion, we have investigated spinon Anderson localization in copper benzoate, an ideal compound of spin-1/2 antiferromagnetic Heisenberg chain. While a linear temperature dependence of spinon specific heat $C_s$ is observed down to 50 mK, the spinon thermal conductivity $\kappa_s$ only shows linear temperature dependence down to 300 mK. Below 300 mK, $\kappa_s/T$ decreases rapidly and vanishes at about 100 mK, which is interpreted as a strong evidence for Anderson localization. We believe that our work is the first example of Anderson localization for magnetic excitations, thus opens a new window for studying such a fundamental phenomenon in condensed matter physics.

**METHODS SUMMARY**

Copper benzoate, $Cu(C_6H_5COO)_2 \cdot 3H_2O$, single crystals were grown from water solutions by a diffusion method[21]. The obtained single crystals are flat plates in the *ac* plane, with the longest dimension up to several centimetres along the chain direction (*c*-axis). Specific heat measurements were carried in a small dilution refrigerator adapted in a Physical Property Measure System (PPMS, Quantum Design). The samples of bar shape were cut and cleaved to typical dimensions $2.0 \times 0.5 \times 0.03$ mm$^3$, with the longest dimension along *c*-axis and the shortest along *b*-axis. Four silver wires were attached on the sample by silver paint. Thermal conductivity were measured in a dilution refrigerator (Oxford Instruments), using a standard four-wire steady-state method with two $RuO_2$ chip thermometers, calibrated in situ against a reference $RuO_2$ thermometer. Magnetic fields were applied along the *b* axis for both specific heat and thermal conductivity measurements.

**Acknowledgements** This work is supported by the Natural Science Foundation of China, the Ministry of Science and Technology of China (National Basic Research Program Nos. 2009CB929203 and 2012CB821402), Program for Professor of Special Appointment (Eastern Scholar) at Shanghai Institutions of Higher Learning.

**Author Information** The authors declare no competing financial interests. Correspondence and requests for materials should be addressed to S. Y. Li (shiyan_li@fudan.edu.cn).

**Figure 1 | Spinons in the spin-1/2 antiferromagnetic Heisenberg chain and crystal structure of copper benzoate.**

**a,** Carton picture of spinon excitations in a spin-1/2 antiferromagnetic Heisenberg chain. **b,** The "two-spinon continuum" spectrum, limited by the lower bound $\varepsilon_1 = \frac{\pi}{2} J |\sin q|$ and upper bound $\varepsilon_2 = \pi J |\sin q/2|$. The spinons are gapless at $q = 0$ and $\pi$. **c,** Crystal structure of copper benzoate. It has a monoclinic crystal structure with lattice constants $a$ = 6.98 Å, $b$ = 34.12 Å, $c$ = 6.30 Å, and $\beta$ = 89.5º. The $Cu^{2+}$ ions carry spin-1/2 and form chains along the $c$ axis. The intrachain coupling $J/k_B \approx$ 18.6 K and the interchain coupling $J'$ is extremely weak, which make copper benzoate an ideal compound of spin-1/2 antiferromagnetic Heisenberg chain.

**Figure 2 | Specific heat of copper benzoate single crystal.**

**a,** Total specific heat of copper benzoate single crystal in zero magnetic field. The solid line is a fit to $C_{total} = aT + bT^3$, in which the first and second terms are contributed by spinons and phonons, respectively. Based on the fitting, the phonons only contribute about 3% of the total specific heat at $T = 1$ K. **b,** Magnetic specific heat in $H = 0$ and 7 T. The small phonon contribution has been subtracted in both cases. The zero-field magnetic thermal conductivity is attributed to spinons. In $H = 7$ T, the thermal conductivity comes from solitons, antisolitons, and breathers, according to the quantum sine-Gordon model. The curve shows a clear signature of energy gap, which can be fitted in the framework of sine-Gordon theory with a gap of 2.41 K.

**Figure 3 | Thermal conductivity of copper benzoate single crystal.**

**a,** $\kappa/T$ vs $T$ in $H = 0$, 7, and 14.5 T for copper benzoate sample S1. The data in $H = 0$ T between 0.3 and 0.7 K can be well fitted by $\kappa = aT + bT^\alpha$, giving $a = 1.70 \pm 0.04$ mW/cm K$^2$, $b = 8.87 \pm 0.04$, and $\alpha = 2.80 \pm 0.03$. The linear term $aT$ is attributed to spinons, and the term $bT^\alpha$ is a typical contribution of phonons. **b,** The magnetic thermal conductivity $\kappa_s$ and specific heat $C_s$ (scaled by 6 times) in $H = 7$ T. The phonon contribution has been subtracted in both cases. $\kappa_s$ shows nearly the same temperature dependence as $C_s$. **c,** $\kappa/T$ vs $T^{1.90}$ for copper benzoate sample S2 in $H = 0$ T. **d,** $\kappa/T$ vs $T^{1.85}$ for copper benzoate sample S3 in $H = 0$ T.

**Figure 4 | Spinon Anderson localization in copper benzoate.**

Spinon thermal conductivity of copper benzoate single crystals S1, S2, and S3 in $H = 0$ T, plotted as $\kappa_s/T$ vs $T$. The different values of $\kappa_s/T$ above 300 mK suggest that the spinon mean free path in copper benzoate is sample dependent. Below 300 mK, $\kappa_s/T$ decreases rapidly and vanishes at about 100 mK, which gives a strong evidence for spinon Anderson localization. The onset temperature of the spinon localization $T_{AL}$ remains at about 300 mK when the mean free path of spinons differs by 3 times between S1 and S2.

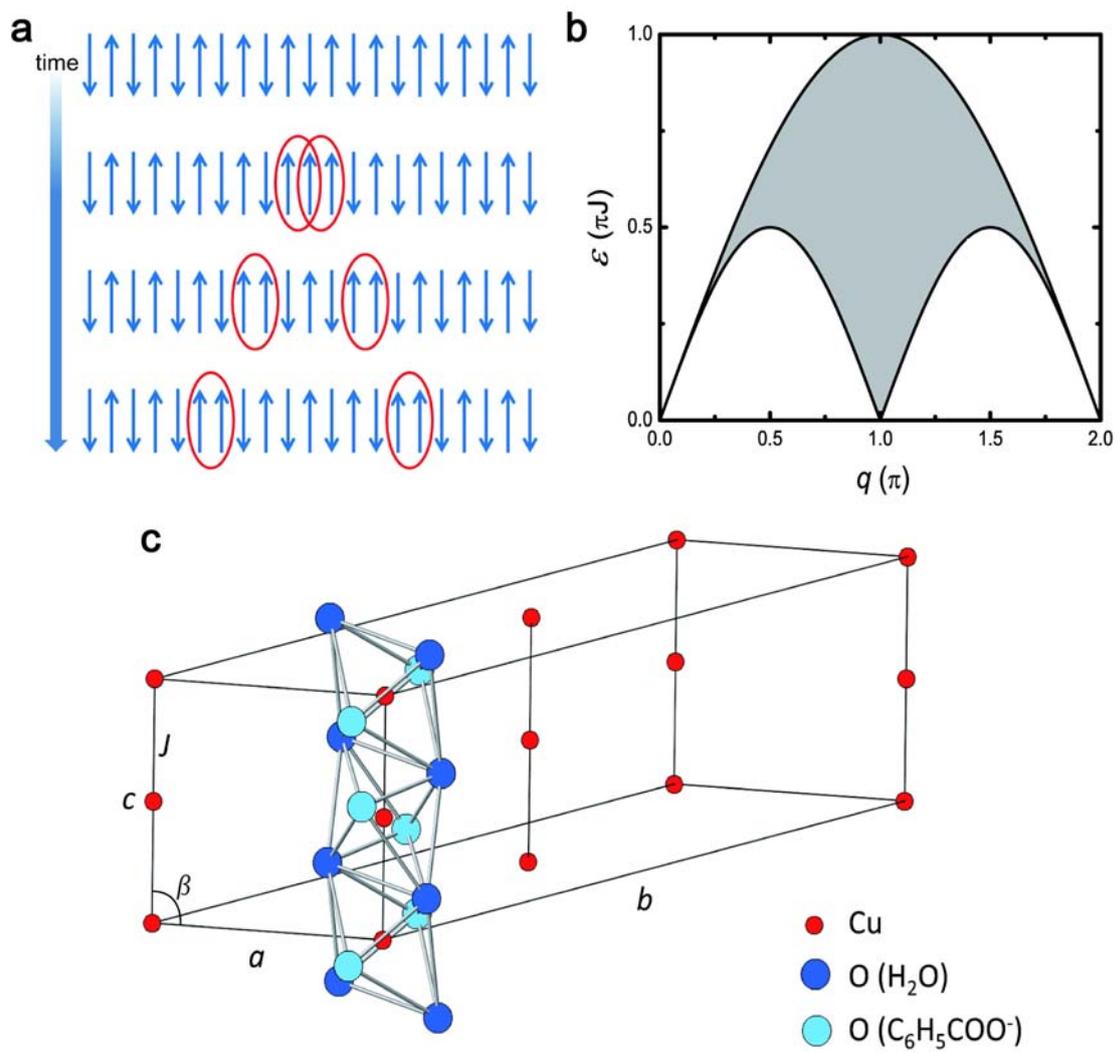

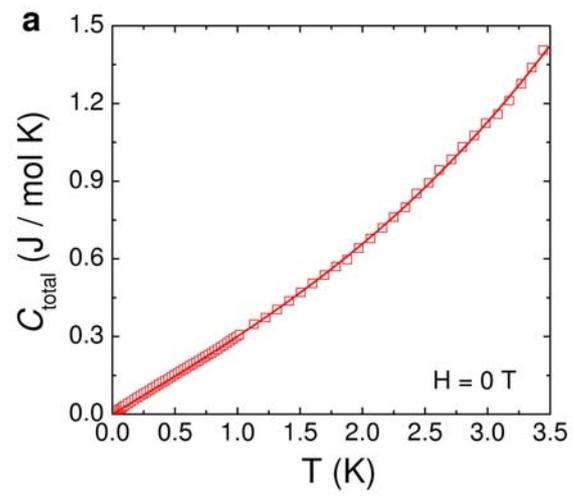

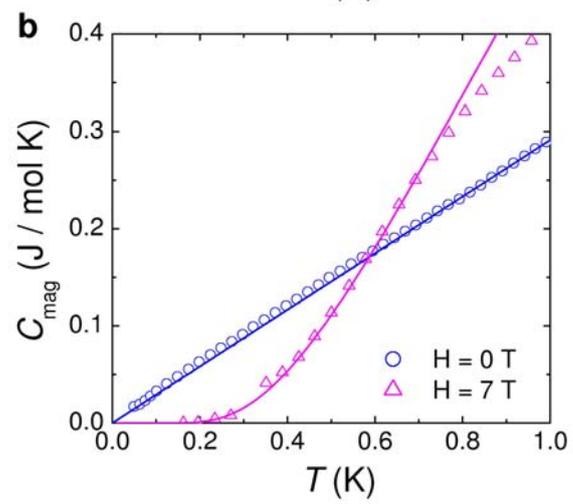

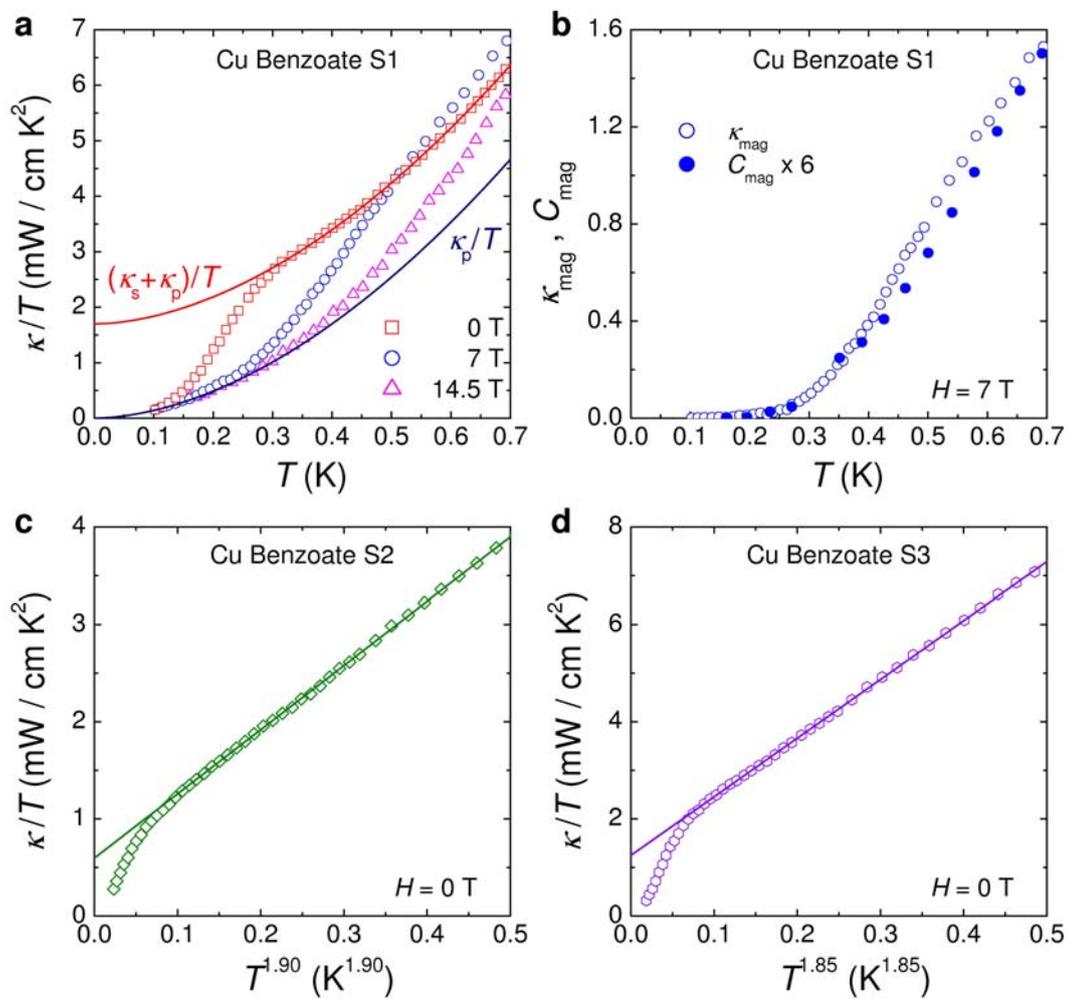

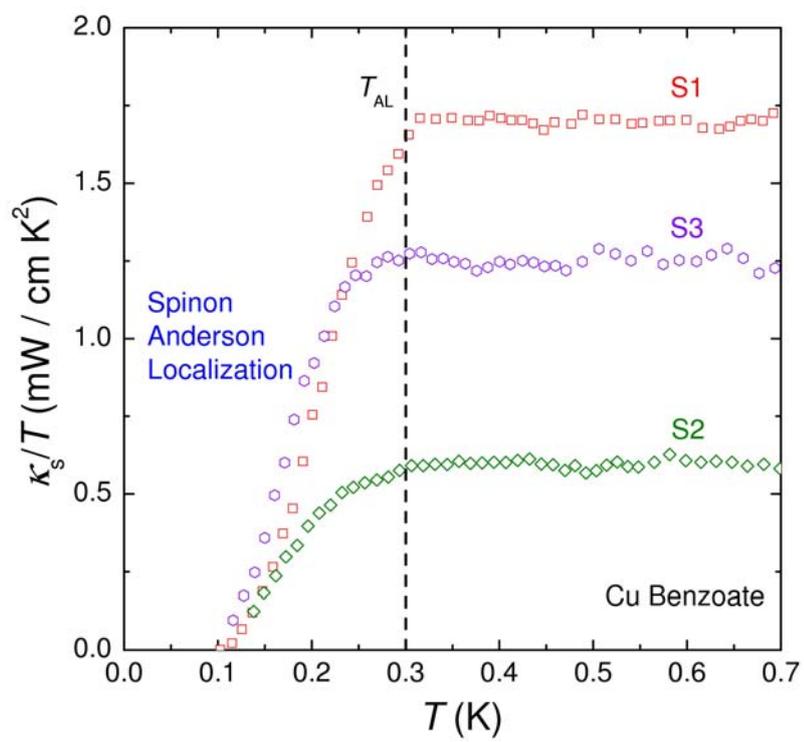